\documentclass[letterpaper,aps,prl,twocolumn,floatfix,amsfonts,amssymb,superscriptaddress]{revtex4-1}
\usepackage{amsmath}
\usepackage{amsthm}
\usepackage{amssymb}
\usepackage{graphicx}
\usepackage{tabularx}

\newcommand{\comment}[1]{}

\begin{document}
\title{Generalized Wilson Chain for solving multichannel quantum impurity problems}

\author{Andrew K. Mitchell}
\affiliation{Department of Chemistry, Physical and Theoretical Chemistry, Oxford University, South Parks Road, Oxford OX1 3QZ, United Kingdom}
\author{Martin R. Galpin}
\affiliation{Department of Chemistry, Physical and Theoretical Chemistry, Oxford University, South Parks Road, Oxford OX1 3QZ, United Kingdom}
\author{Samuel Wilson-Fletcher}
\affiliation{Department of Chemistry, Physical and Theoretical Chemistry, Oxford University, South Parks Road, Oxford OX1 3QZ, United Kingdom}
\author{David E. Logan}
\affiliation{Department of Chemistry, Physical and Theoretical Chemistry, Oxford University, South Parks Road, Oxford OX1 3QZ, United Kingdom}
\author{Ralf Bulla}
\affiliation{Institute for Theoretical Physics, University of Cologne, 50937 Cologne, Germany}


\begin{abstract}
The Numerical Renormalization Group is used to solve quantum impurity problems, which describe magnetic impurities in metals, nanodevices, and correlated materials within DMFT. 
Here we present a simple generalization of the Wilson Chain, which improves the scaling of computational cost with the number of channels/bands, bringing new problems within reach. 
The method is applied to calculate the t-matrix of the three-channel Kondo model at $T=0$, which shows universal crossovers near non-Fermi liquid critical points. A non-integrable three-impurity problem with three bands is also studied, revealing a rich phase diagram and novel screening/overscreening mechanisms.
\end{abstract}

\maketitle


Wilson's Numerical Renormalization Group (NRG) has been widely used over the last 40 years to study systems where local interacting degrees of freedom are coupled to bands (or channels) of non-interacting conduction electrons \cite{hewson,wilson,nrg:rev}. 
Such `quantum impurity problems' are classic paradigms for strong electron correlations in condensed matter. They appear in diverse contexts, being fundamental to the theoretical description of both nanostructures \cite{kondo_revival,1ckexpta, *1ckexptb,liang,nygard,akm:MF} and correlated materials \cite{hewson,colemanhf,sihf}.

NRG calculations have so far been largely restricted to problems involving one or two conduction bands \cite{nrg:rev}. Although more complicated systems sometimes reduce to an effective single-channel description \cite{hewson,wilson,nrg:rev,kondo_revival,1ckexpta, *1ckexptb}, many real systems of interest involve three or more bands. For example, transition metal oxides such as LaTiO$_3$, SrVO$_3$ and SrRuO$_3$ are described  by the `Hubbard-Kanamori' model of interacting $t_{2g}$ orbitals \cite{mit_rev, georgescorr}, which maps onto an effective three band problem involving three orbitals within dynamical mean field theory (DMFT) \cite{dmft_rev}. Likewise, treatment of two-dimensional systems such as the cuprates within cluster DMFT necessitates solution of multichannel coupled impurity models \cite{DCA_dmft}. Indeed, interest in quantum impurities was originally stimulated by the Kondo effect due to iron impurities in gold, described quantitatively by a spin-$\tfrac{3}{2}$ impurity, exactly-screened by three conduction bands \cite{FeinAu,*FeinAuII}. 

The major advantage of NRG over other numerically-controlled `impurity solvers', such as exact diagonalization \cite{ED} and continuous-time quantum Monte Carlo \cite{CTQMC}, is that it yields highly accurate results on essentially all energy/temperature scales \cite{note_resolution}, for very general impurity problems with arbitrary interactions. However, a serious limitation of NRG is that calculational cost scales exponentially with the number of channels. In practice, this has largely limited application of NRG to two-channel problems \cite{nrg:rev} and special high-symmetry systems \cite{nonabsym}.

In this work we develop a generalized Wilson chain, which significantly reduces the computational cost of solving multichannel models with NRG. The effective single-channel formulation brings a new range of more complex multichannel problems within reach of NRG. The generic class of Hamiltonian considered is $H = H^0 + \sum_{\alpha=1}^{N_c} H_{CB}^{\alpha}$, with each of the $N_c$ conduction bands described by
\begin{eqnarray}
\label{eq:H_CB}
 H_{CB}^{\alpha} = \sum_{k,\sigma} \epsilon^{\phantom{\dagger}}_{\alpha k} c^{\dagger}_{\alpha k\sigma}c^{\phantom{\dagger}}_{\alpha k\sigma} \;,
\end{eqnarray}
and $H^0$ describing the interacting `impurity' subsystem, coupled locally to these bands. The technique is general; exemplified here by the multichannel Kondo (MCK) \cite{nozieres} and multi-impurity Kondo (MIK) \cite{jones} models,
\begin{eqnarray}
\label{eq:H_MCK}
 H^0_{MCK} &=& \sum_{\alpha} J_{\alpha}\hat{\textbf{S}}_1\cdot   \hat{\textbf{s}}_{\alpha} \;, \\
\label{eq:H_MIK}
 H^0_{MIK} &=& \sum_{\alpha} J_{\alpha}\hat{\textbf{S}}_{\alpha}\cdot \hat{\textbf{s}}_{\alpha}  + K \sum_{\alpha<\alpha'} \hat{\textbf{S}}_{\alpha}\cdot  \hat{\textbf{S}}_{\alpha'} \;,
\end{eqnarray}
where $\hat{\textbf{s}}_{\alpha} = \sum_{\sigma,\sigma'} \tfrac{1}{2} f^{\dagger}_{\alpha 0 \sigma} \vec{\sigma}_{\sigma,\sigma'} f^{\phantom{\dagger}}_{\alpha 0 \sigma'}$ is the electronic spin density in channel $\alpha$ (here $f_{\alpha 0 \sigma} = N_\text{orb}^{-1/2}\sum_k c_{\alpha k \sigma}$, and $N_\text{orb}\rightarrow \infty$), and  $\hat{\textbf{S}}_{i}$ are impurity spin-$\tfrac{1}{2}$ operators. 

\begin{figure}[h!]
\begin{center}
\includegraphics[width=61mm]{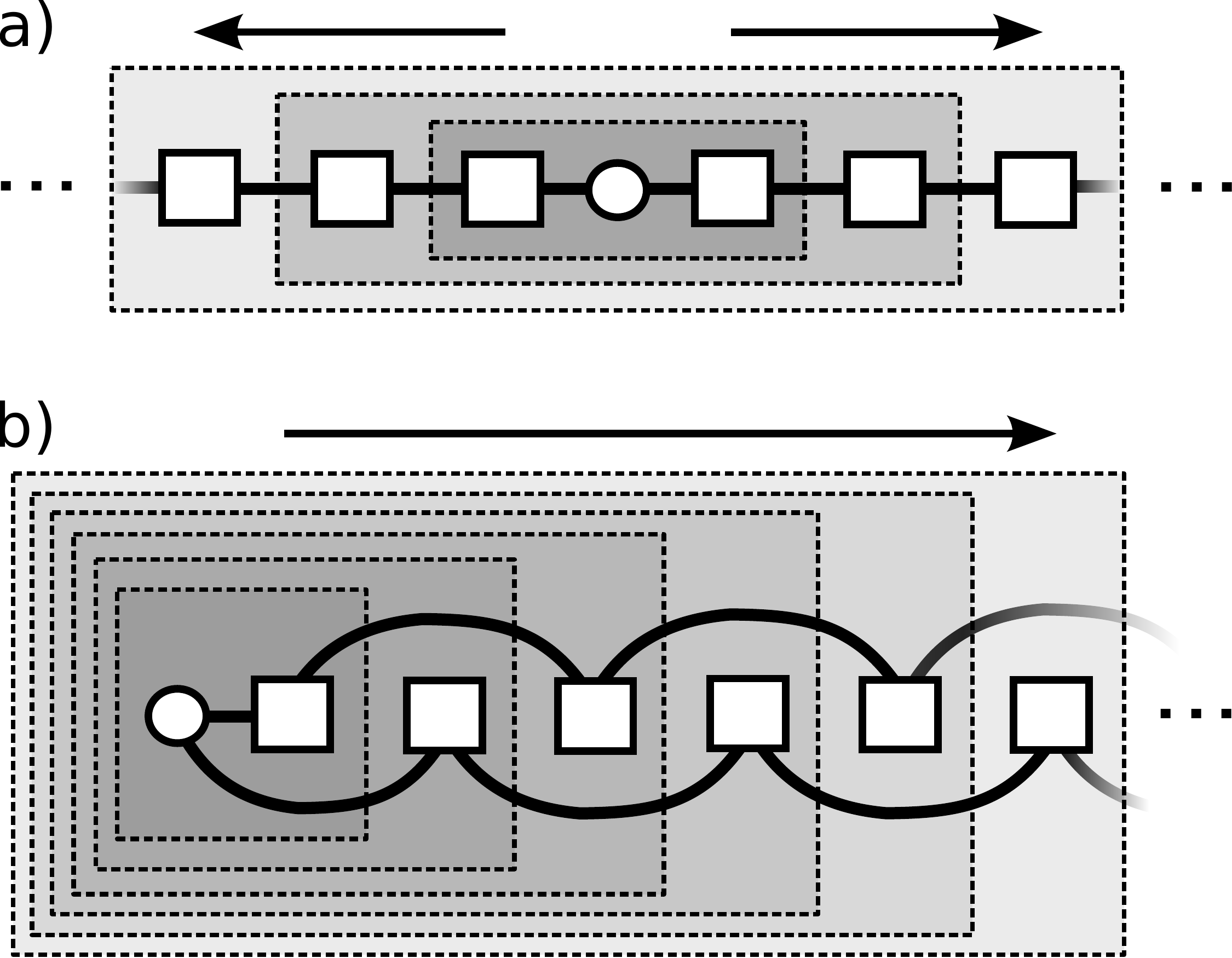}
\caption{\label{fig:wc}
NRG for a two-channel model. Impurity subsystem and Wilson chain orbitals denoted by the circle and squares, respectively. Solid lines represent chain couplings; dashed boxes represent diagonalization and truncation. Energy scale decreases in the direction of the arrows. (a) Standard formulation: channel orbitals added symmetrically. (b) Interleaved formulation: channel orbitals added sequentially.
}
\end{center}
\end{figure}

We focus on $N_c=3$ band problems, calculating previously inaccessible dynamics of the 3CK model at $T=0$. The universal crossover to the non-Fermi liquid 3CK quantum critical point (QCP) 
is extracted, as are subsequent non-trivial crossovers to 2CK or Fermi liquid states induced by a single detuning perturbation. The non-integrable 3IK model is also studied, with a rich range of physics associated with simpler 2CK, 3CK and 2IK models clearly manifest in thermodynamic quantities.


\emph{Numerical Renormalization Group.--} 
The key approximation of NRG is a \emph{logarithmic} discretization of the conduction electron densities of states, $\rho_\alpha(E)$, which captures  low-energy excitations around the Fermi level crucial to Kondo physics \cite{wilson,nrg:rev}. For simplicity, we take 
standard
symmetric bands defined between  $E=\pm D_\alpha$, with constant density of states $\rho_\alpha = 1/(2D_\alpha)$. Discretization points for each band at $E= \pm x_{n}$ are given by,
\begin{equation}
\label{eq:discpoints}
x_{n} = \begin{cases} D_{\alpha} & :~~n=0 \\ 
  D_{\alpha} \Lambda^{-n-z_{\alpha}} & :~~n=1,2,3, ... \end{cases} 
\end{equation}
with $\Lambda>1$ and $z_{\alpha} \ge 0$ \cite{wilson,nrg:rev,ztrick}. Retaining only the symmetric linear combinations of states in each interval  of a given conduction channel allows its Hamiltonian to be written in the form of a `Wilson chain': 
\begin{equation}
\label{eq:wc}
H^{\alpha,\textit{disc}}_{CB} = \sum_{\sigma} \sum_{n=0}^{\infty}\left ( t^{\phantom{\dagger}}_{\alpha n} f^{\dagger}_{\alpha n \sigma}f_{\alpha (n+1) \sigma} + \text{H.c.} \right ) \;,
\end{equation}
with hopping matrix elements $t_{\alpha n}$ \cite{wilson, nrg:rev}. 
The (discretized) MCK or MIK Hamiltonian is then built up iteratively, as shown 
in Fig.~\ref{fig:wc}(a) for $N_c=2$. The first step (innermost gray box) corresponds to the impurity-$f_{\alpha 0 \sigma}$ subsystem, given by $H^0$ [Eqs.~\ref{eq:H_MCK} or \ref{eq:H_MIK}]. 
A complete `shell' of $N_c$ Wilson chain sites is then added symmetrically \cite{[{An asymmetric truncation scheme was used in }] asymtrunc} at each iteration using the recursion relation \cite{wilson,nrg:rev},
\begin{eqnarray}
\label{eq:HN_rec}
H^{N+1} = H^{N}  + \sum_{\alpha,\sigma}\left ( t^{\phantom{\dagger}}_{\alpha N} f^{\dagger}_{\alpha N \sigma}f^{\phantom{\dagger}}_{\alpha (N+1) \sigma} + \text{H.c.} \right ),
\end{eqnarray}
with $H^{N+1}$ rediagonalized after each shell is added, and the full Hamiltonian recovered as $H^{\textit{disc}}=  \lim_{N\rightarrow \infty} H^{N}$.

The dimension of the Hilbert space thus grows by a factor of $4^{N_c}$
at each iteration. In practice, the calculation is made 
tractable by retaining a large but \emph{fixed}
number, $N_s$, of the lowest-energy states at each iteration. This
is justified by a special
property of the Wilson chain hoppings, which decrease exponentially
down each chain due to the logarithmic discretization
\cite{wilson,nrg:rev,ztrick}. After a few iterations (such that
$\Lambda^{-n/2}\ll 1$),
\begin{equation}
\label{eq:hop}
t_{\alpha n} \sim D_{\alpha} \Lambda^{-z_{\alpha}} \times \Lambda^{-n/2} \;.
\end{equation}
This leads to an \emph{energy scale separation} from one iteration to
the next. The high-energy eigenstates of $H^N$ can thus be discarded
without affecting the low-energy states of $H^{N+1}$, and
numerically-exact results are obtained \cite{wilson}.

But at each iteration, after adding $N_c$ new Wilson chain orbitals, a Hilbert space of total dimension $N_t = N_s\times 4^{N_c}$ must still be diagonalized: this is the computational bottleneck~\cite{sm}.
Although symmetries may be exploited to reduce the total cost \cite{wilson,nrg:rev,nonabsym}, this exponential scaling of $N_t$ with the number of channels currently prevents standard NRG from being used to calculate dynamics for generic models with $N_c>2$.


\emph{Interleaved Wilson chains.--} 
Here we show that the scaling of standard NRG with the number of channels can be significantly improved upon by introducing additional energy scale separations \emph{between channels}. 
All $N_c$ Wilson chains are then \emph{interleaved} to form a single generalized Wilson chain, 
depicted in Fig.~\ref{fig:wc} (b). Importantly, the Hilbert space can then be truncated after each orbital is added \emph{in sequence}. 
Although this procedure requires $N_c$ more matrix diagonalizations, the Hilbert space at any given step is much smaller than in standard NRG ($N_t = N_s\times 4$, rather than $N_s\times 4^{N_c}$). This effective single-channel formulation significantly reduces the overall cost of the calculation \cite{sm}.
Denoting by $H^{N+1}_{M}$ the Hamiltonian in which $M$ of the $N_{c}$ orbitals in shell $N+1$ have been added, the recursion relation Eq.~(\ref{eq:HN_rec}) generalizes to 
\begin{eqnarray}
\label{eq:HN_rec2}
H^{N+1}_{M} = H^{N+1}_{M-1} +\sum_{\sigma}\left ( t^{\phantom{\dagger}}_{M N} f^{\dagger}_{MN \sigma}f^{\phantom{\dagger}}_{M(N+1) \sigma} + \text{H.c.} \right ), \qquad
\end{eqnarray}
(where $H^{N+1}_0 = H^{N}_{N_c}$ $\equiv  H^{N}$). Each $H^{N+1}_{M}$ is diagonalized and the Hilbert space truncated down to $N_s$ states before moving to $H^{N+1}_{M+1}$.  Truncation is possible if an energy scale separation exists between each step $M\rightarrow M+1$. This arises automatically if the channel bandwidths satisfy $D_{M+1}/ D_{M} = g = \Lambda^{-1/(2N_c)}$. It can also be introduced artificially at the discretization step by choosing slide parameters $z_{M+1} - z_{M} = 1/2N_c$. In each case it then follows from Eq.~(\ref{eq:hop}) that $t_{(M+1)N}/t_{MN} = g$. We present results using both methods below. 

Note that in standard NRG, the energy scale decreases by $\Lambda^{-1/2}$ at each iteration, while the factor is $g=\Lambda^{-1/(2N_c)}$ for the interleaved method. A larger $\Lambda$ is therefore needed for $N_c>1$ to justify truncation. This also has the effect of amplifying discretization artefacts \cite{wilson}, but these can be diminished \cite{note_resolution} by averaging over band discretizations using the slide parameter $z_{\alpha}$. 

By complete analogy to standard NRG, thermodynamics at an effective temperature $T\sim t_{MN}$ can be calculated at each step from the eigenstates of $H^{N}_{M}$ \cite{wilson}, and dynamics can be calculated from the full density matrix \cite{fdmnrg,*asbasis} in the Anders-Schiller basis \cite{asbasisprl,note_as}.


\begin{figure*}[t]
\begin{center}
\includegraphics[width=180mm]{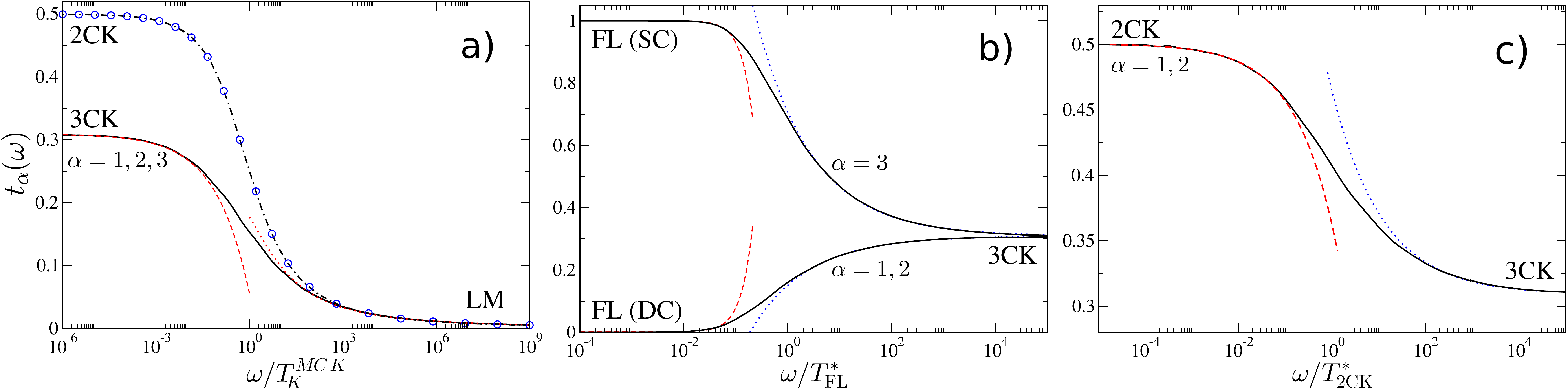}
\caption{\label{fig:mck}
Scaling spectra $t_{\alpha}(\omega)$ for channel $\alpha=1,2,3$ at $T=0$ for the 3CK model, obtained using the interleaved NRG method. 
For parameters, see text. T-matrices for equivalent channels have been averaged \cite{sm}. 
(a) vs $\omega/T_K^{3CK}$, tuned to the critical point with $\rho_1 J^c_1=0.05$, $\rho_2 J^c_2\simeq 0.05088...$, $\rho_3 J^c_3\simeq 0.05180...$ such that $T_K^{3CK}\approx 10^{-10} D_1$. Dot-dashed line is for the 2CK model, with points obtained using standard NRG for equal bandwidths ($g=1$). (b) Crossover to FL fixed points due to perturbation $J_3=J_3^c+0.01\times T_K^{3CK}$. The impurity flows to strong coupling [FL (SC)] with channel 3, while channels 1 and 2 decouple [FL (DC)]. (c) Crossover to 2CK fixed point due to perturbation $J_3=J_3^c-0.01\times T_K^{3CK}$. Channel 3 decouples, while the impurity is overscreened by channels 1 and 2. Asymptotes for (a)--(c) given by Eqs.~(\ref{eq:tm_3ck})--(\ref{eq:tm_2ck}), and discussed further in the text.
}
\end{center}
\end{figure*}

\emph{Multichannel Kondo.--}
We present results for the 3CK model [Eqs.~\ref{eq:H_CB} and \ref{eq:H_MCK} with $N_c=3$], a stringent test of our method since the critical physics is destabilized by various symmetry-breaking perturbations \cite{nozieres,cragg_lloyd,CFT2CK,andreidestri,*tsvelik,potok,akm:exactNFL,*akm:finiteT}. The interleaved NRG method provides access to dynamical quantities not previously calculable within standard NRG. We focus on the $T=0$ scattering t-matrix $\mathcal{T}^{k,k'}_{\alpha}(\omega)$ for channel $\alpha=1,2,3$ \cite{hewson,akm:oddimp}.

The model is taken with common bandwidth ratio $D_2/D_1=D_3/D_2=g$, where $g=\Lambda^{-1/6}$ as above (and $z_i\equiv z$ identical for all channels). 
We use $\Lambda=10$, and average the results of three NRG runs with $z=\tfrac{1}{6},\tfrac{1}{2},\tfrac{5}{6}$. Although $N_s=20000$ is required here, intermediate truncations within interleaved NRG mean that each run can be performed on a standard desktop computer in a few hours (a factor of $\sim 600$ faster than standard NRG \cite{sm}).

The MCK critical physics occurs at the point of frustration where all channels couple equally to the impurity and no single channel can completely screen it. 
For wide flat-bands $D_{\alpha}\rightarrow \infty$ (with finite $g$), the coupling of each channel to the impurity is characterized by the dimensionless quantity $\rho_\alpha J_\alpha$, and criticality arises when $J_2/J_1 = J_3/J_2 = g$. The same critical physics arises for finite bandwidths, with the position of the critical point itself naturally shifted slightly \cite{sm}. Discretization and asymmetric truncation in NRG produces a further tiny deviation in the critical ratios $J_{\alpha}/J_{\beta}$ \cite{sm,note_tuning}. 

As exemplified by Fig.~\ref{fig:mck} (a), the scaling spectrum of the t-matrix, $t_{\alpha}(\omega)=-N_{\text{orb}} \pi\rho_{\alpha} ~\text{Im}\mathcal{T}^{k,k'}_{\alpha}(\omega)$ is captured by interleaved NRG at criticality (plotted vs $\omega/T_K^{3CK}$ [solid line], with $T_K^{3CK}$ the 3CK Kondo scale). In the universal regime $|\omega| \ll D_{\alpha}$, $t_{\alpha}(\omega)$ is identical for all three channels \cite{sm}, establishing the emergent channel symmetry. For $|\omega|\gg T_K^{3CK}$ (dotted line) the physics is controlled by spin-flip scattering near the LM fixed point --- behavior common to all quantum impurity models where local moment physics plays a key role \cite{LMA_scalingspec}. By contrast, for $|\omega|\ll T_K^{3CK}$ (dashed line), the physics is controlled by corrections to the 3CK fixed point. We find,
\begin{equation}
\label{eq:tm_3ck}
\hskip -0.08cm t_{\alpha}(\omega) = \begin{cases}
    a/[b+c\ln^2(|\omega|/T_K^{3CK})]  &  : ~~|\omega|\gg T_K^{3CK}  ,\\
    \gamma-d(|\omega|/T_K^{3CK})^{2/5}          &  : ~~|\omega|\ll T_K^{3CK}  ,
  \end{cases}
\end{equation}
where the  $2/5$ power arises from the scaling dimension~of the leading irrelevant operator at the 3CK fixed point~and $\gamma=\cos(2\pi/5) \simeq 0.31$ \cite{CFT2CK}. The result using interleaved NRG for the 2CK model is shown as the dot-dashed line for comparison [and agrees perfectly with standard NRG for the \emph{equal bandwidth} case $g=1$ (circles)].

Fig.~\ref{fig:mck} (b) shows the scaling spectra $t_{\alpha}(\omega)$ vs $\omega/T_{\textit{FL}}^*$ when $(J_3-J_3^c)>0$ ($\ll T_K^{3CK}$).
We find this generates a universal crossover from the 3CK fixed point to a Fermi liquid (FL) fixed point on the scale of $T_{\textit{FL}}^*\propto T_K^{3CK}|\rho_3 J_3-\rho_3 J_3^c|^{5/2}$. The t-matrix for channel 3 shows a Kondo resonance, with $t_{3}(0) =1$ signalling single-channel strong coupling physics; while channels 1 and 2 decouple asymptotically,  yielding $t_{1,2}(0)=0$. 
We find the asymptotic behavior to be,
\begin{equation}
\label{eq:tm_fl}
t_{\alpha}(\omega) = \begin{cases}
   \gamma_{\phantom{\alpha}} + e_{\alpha}(|\omega|/T_{FL}^*)^{-2/5}  & : ~~|\omega|\gg T_{FL}^{*}  ,\\
   \xi_{\alpha} - f_{\alpha}(|\omega|/T_{FL}^*)^2             & : ~~|\omega|\ll T_{FL}^{*}  ,
  \end{cases}
\end{equation}
where $\xi_{\alpha}=t_{\alpha}(0)=0$ or $1$ characteristic of the FL fixed point as above; and $e_{\alpha}$, $f_{\alpha} >0$ for channel $\alpha=3$ and $<0$ for $\alpha=1,2$. 
Strikingly, when the sign of the perturbation is simply reversed, $(J_3-J_3^c)<0$, channel 3 decouples [yielding the same scaling curve as $t_{1,2}(\omega)$ in Fig.~\ref{fig:mck} (b)], but the impurity spin is overscreened on the scale of $T_{2CK}^* \propto T_K^{3CK}|\rho_3 J_3-\rho_3 J_3^c|^{5/2}$ by channels 1 and 2, producing stable 2CK physics characterized by, 
\begin{equation}
\label{eq:tm_2ck}
\hskip -0.02cm t_{1,2}(\omega) = \begin{cases}
   \gamma + g(|\omega|/T_{2CK}^*)^{-2/5}         & : ~~|\omega|\gg T_{2CK}^{*}  ,\\
   \tfrac{1}{2} - h(|\omega|/T_{2CK}^*)^{1/2}  &  : ~~|\omega|\ll T_{2CK}^{*}  .
  \end{cases}
\end{equation}
Eqs.~(\ref{eq:tm_3ck})--(\ref{eq:tm_2ck}) are  consistent with CFT results \cite{note_eran}, and all coefficients $\{a,...,h\}=\mathcal{O}(1)$.


\begin{figure}[t]
\begin{center}
\includegraphics[width=75mm]{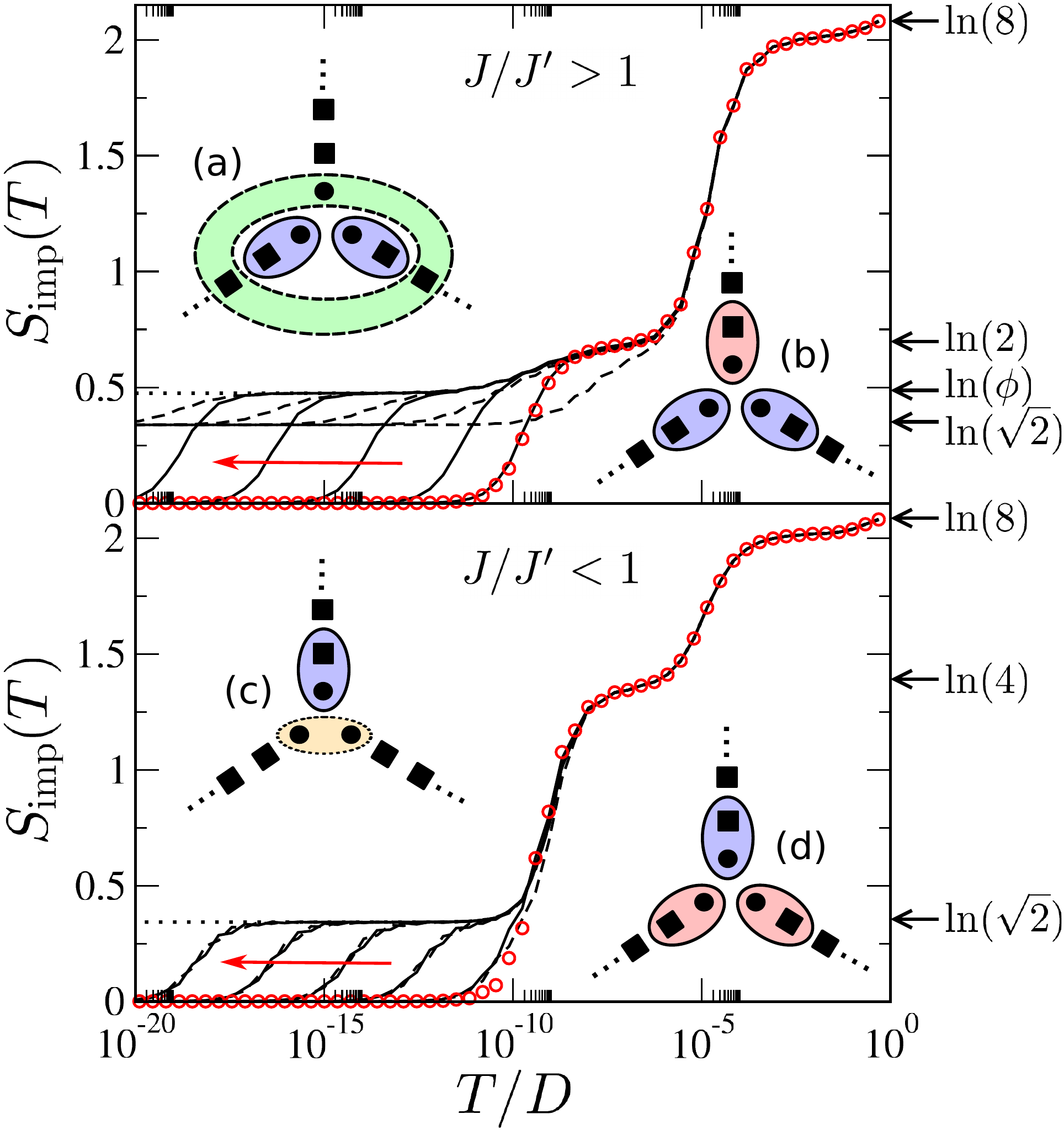}
\caption{\label{fig:3ik}
Impurity contribution to entropy $S_{\text{imp}}(T)$ vs $T/D$ for the 3IK model. Upper panel: $J/D=0.2$, $J'/D=0.1$, tuning $K=K_c\pm \lambda T_K^c$ to approach the 3CK QCP (dotted line) from the 2CK phase ($K>K_c$, dashed lines) and from the Kondo-screened FL phase ($K<K_c$, solid lines). At the QCP, $K_c\simeq 7\times 10^{-6}D$ and $T_K^c\approx 10^{-10} D$. Plotted for $\lambda= 10^{4}, 10^{3}, 10^{2}, 10, 1$ in the direction of the arrow. Circle points for $K=0$. 
Lower panel: analogous plot for $J/D=0.1$, $J'/D=0.2$. The 2IK QCP with $K_c\simeq 10^{-9}D$, $T_K^c\approx 10^{-9} D$ separates local singlet and Kondo screened phases. Plotted for $\lambda= 1, 10^{-1}, 10^{-2}, 10^{-3}, 10^{-4}$ in the direction of the arrow. 
(a)--(d) illustrate the ground states on each side of the transition (see text), using solid ellipses for Kondo singlets, dashed ellipse for 2CK overscreening, dotted ellipse for local singlet. 
}
\end{center}
\end{figure}

\emph{Multi-Impurity Kondo.--} 
The 2IK model [Eqs.~\ref{eq:H_CB} and \ref{eq:H_MIK} with $N_c=2$] describes two exchange-coupled impurities, each coupled also to its own conduction band. It has a non-Fermi liquid QCP separating local singlet and Kondo-screened phases \cite{jones,CFT2IKM,akm:exactNFL,*akm:finiteT,akm:2ckin2ik}. Here we consider the $N_c=3$ (3IK) generalization, focussing on the parity-symmetric case, $J_{1,2}\equiv J$ and $J_3 \equiv J'$, with common bandwidths $D_{\alpha}\equiv D$. To achieve energy-scale separation between channels we use $\Lambda=5$ with $z_1=z_2=0$ and $z_3=\tfrac{1}{4}$, retaining $N_s=10000$ states after chain orbitals of channels 2 and 3 have been added, but keeping \emph{all} states after adding an orbital to channel 1. This hybrid method has the advantage that exact parity symmetry is preserved in the calculation, so tuning is not required to enforce this symmetry \cite{note_calctime}.

We show results for two cases: $J > J'$ and $J<J'$. In each case, a quantum phase transition arises at a critical interimpurity coupling, $K=K_c$. These transitions are distinct \cite{nutshell} and described separately below.

In Fig.~\ref{fig:3ik} (upper panel) we take $J/J'=2$ and show the impurity contribution to the entropy $S_{\text{imp}}(T)$, versus $T/D$.  The circle points are for $K=0$: here each impurity is clearly Kondo screened by its own channel [illustrated in (b)], appearing overall as a two-stage process since $J\ne J'$. On initially increasing $K$ (solid lines, following the arrow), the NRG flows approach progressively more closely the QCP, before ultimately crossing over to the fully Kondo-screened ground state, with $S_{\text{imp}}(0)=0$. At $K=K_c$ (dotted line) the crossover scale vanishes and the QCP is stable. On increasing $K$ further (dashed lines, against the direction of the arrow) impurity 3 feels an \emph{effective} coupling to the remaining Fermi liquid bath degrees of freedom in channels 1 and 2, mediated via the preformed Kondo singlets \cite{akm:oddimp,akm:2ckin2ik,akm:TQD1ch_tom}. This effective coupling becomes stronger than the coupling of impurity 3 to its own channel for $K>K_c$, resulting in its 2CK \emph{overscreening} by symmetric coupling to channels 1 and 2 [illustrated in (a)], generating the residual 2CK entropy $\tfrac{1}{2}\ln(2)$ \cite{andreidestri,*tsvelik,akm:tqd2ch}. The QCP itself can be understood as the point of frustration where the effective coupling to all three channels is equal, resulting in 3CK physics, with entropy $\ln(\phi)$~\cite{andreidestri,*tsvelik} 
(and $\phi=\tfrac{1}{2}(1+\sqrt{5})$ the golden ratio).

Fig.~\ref{fig:3ik} (lower panel) shows the behaviour for $J/J'=\tfrac{1}{2}$ when $K$ is similarly increased. For small interimpurity coupling $K<K_c$, the Kondo-screened phase again arises [see (d)]. But for $K>K_c$, while  channel 3 again screens its own impurity first, the other two 
impurities now form a `local' singlet, with overall $S_\mathrm{imp}(0)=0$ [illustrated in (c)]. The competition between Kondo-screened and local-singlet states is the origin of criticality in the 2IK model \cite{jones,CFT2IKM,akm:2ckin2ik} --- and similarly here the QCP is of 2IK type, with characteristic residual entropy $S_{\text{imp}}(0)=\tfrac{1}{2}\ln(2)$ (dotted line). 
Dynamics provide more detailed information about the screening/overscreening mechanisms, and confirm this physical picture. We conjecture that MIK models support all $N \le M$-channel Kondo critical points, by analogous mechanisms.


\emph{Conclusion.--} 
We have described a conceptually and technically simple new method which significantly reduces the cost of studying multichannel models by NRG. An energy-scale separation between channels is introduced, justifying additional Hilbert space truncation. The usual scaling of standard NRG with the number of channels is thereby drastically improved. The scale separation appears automatically when conduction electron channels have different bandwidths, or can be introduced at the discretization step by exploiting the `$z$-trick'~\cite{ztrick}.

The interleaved method allows previously inaccessible calculations to be performed with NRG. This was demonstrated by application to the three-channel and three-impurity Kondo models near their non-Fermi liquid QCPs, for which non-trivial new results were presented. Further applications may include multichannel quantum impurity problems that appear as effective models within DMFT for multi-band correlated lattice problems.


\emph{Acknowledgments.--} We thank E. Sela for insightful discussions, and acknowledge financial support from EPSRC through EP/I032487/1 (AKM, DEL) and the DFG through SFB608 and FOR960 (RB). 

\clearpage


%


\begin{thebibliography}{47}
\bibitem [{not({\natexlab{b}})}]{note_as}%
  \BibitemOpen
  \bibinfo {note} {The Anders-Schiller basis for interleaved NRG comprises all
  discarded states at each step. The environmental degeneracy of eigenstates is
  thus $4^{N_c(N_\text{max}+1-N)-M}$, with $N_\text{max}+1$ the total number of
  sites in each Wilson chain.}\BibitemShut {Stop}%
\bibitem [{not({\natexlab{c}})}]{note_tuning}%
  \BibitemOpen
  \bibinfo {note} {Calculation of critical properties first requires the
  critical point itself to be located. Efficient optimization routines can be
  used to minimize the number of NRG runs required to tune the ratios
  $J_{\alpha}/J_{\beta}$ (dynamics need not be calculated during this
  process).}\BibitemShut {Stop}%
\bibitem [{not({\natexlab{d}})}]{note_calctime}%
  \BibitemOpen
  \bibinfo {note} {The total computational cost is comparable to that of a
  conventional two-channel NRG calculation: thermodynamics as in
  Fig.~\ref{fig:3ik} are obtained in less than 2 hours on a standard desktop
  computer.}\BibitemShut {Stop}%
\end{thebibliography}
\end{document}


\title{Generalized Wilson Chain for solution of multichannel quantum impurity problems:\\ Supplementary Material}

\author{Andrew K. Mitchell}
\affiliation{Department of Chemistry, Physical and Theoretical Chemistry, Oxford University, South Parks Road, Oxford OX1 3QZ, United Kingdom}
\author{Martin R. Galpin}
\affiliation{Department of Chemistry, Physical and Theoretical Chemistry, Oxford University, South Parks Road, Oxford OX1 3QZ, United Kingdom}
\author{Samuel Wilson-Fletcher}
\affiliation{Department of Chemistry, Physical and Theoretical Chemistry, Oxford University, South Parks Road, Oxford OX1 3QZ, United Kingdom}
\author{David E. Logan}
\affiliation{Department of Chemistry, Physical and Theoretical Chemistry, Oxford University, South Parks Road, Oxford OX1 3QZ, United Kingdom}
\author{Ralf Bulla}
\affiliation{Institute for Theoretical Physics, University of Cologne, 50937 Cologne, Germany}


\maketitle


Formulating multichannel quantum impurity problems in terms of a single interleaved Wilson chain allows the solution by NRG to be achieved with a significant reduction in computational cost, as compared with standard NRG. This stems from the fact that the size of the Hilbert space can be truncated at intermediate stages in the calculation. The computational bottleneck within standard NRG --- the need to diagonalize large matrices --- is thus substantially alleviated.  

Solving a particular multichannel problem using the new method requires less computer memory and less CPU time compared to standard NRG. Details of the gain in efficiency are presented in Sec.~I, below. Using present computational resources, accurate solution of a new range of three-channel problems is now possible (as demonstrated for 3CK and 3IK models in the main paper). Indeed, such calculations can be performed on a standard desktop computer in a few hours.


As shown in Sec.~II, adding the complete Wilson `shell' of $N_c$ orbitals simultaneously is not necessary: the eventual error introduced by intermediate truncation is small and does not affect calculated physical properties. In particular, the universal physics is fully captured by the interleaved method due to an emergent channel symmetry, as discussed in Sec.~III.


\section{Efficiency of the Interleaved method}
We consider here as a concrete non-trivial example the spin-$\tfrac{1}{2}$ multichannel Kondo (MCK) model, with $N_c=1,2,3$ and $4$ channels. For generality, we use here a basis of  $U(1)$ conserved charge in each channel $Q_{\alpha}$, and overall conserved spin projection $S^z_{\text{tot}}$. (In the absence of potential scattering or magnetic field, $SU(2)$ spin and isospin symmetries can instead be exploited to block diagonalize NRG Hamiltonians in multiplet space.)

In standard NRG, the computational complexity of diagonalizing the MCK model increases with the number of channels. This is due to a proliferation of low-energy states, corresponding to combinations of low-energy excitations in each of the channels. Since the method works by discarding only the high-energy states at each iteration, the total number of states, $N_s$, which one needs to keep at each iteration to obtain results of similar accuracy thus increases with the number of channels.

\begin{figure}[H]
\begin{center}
\includegraphics[width=63mm]{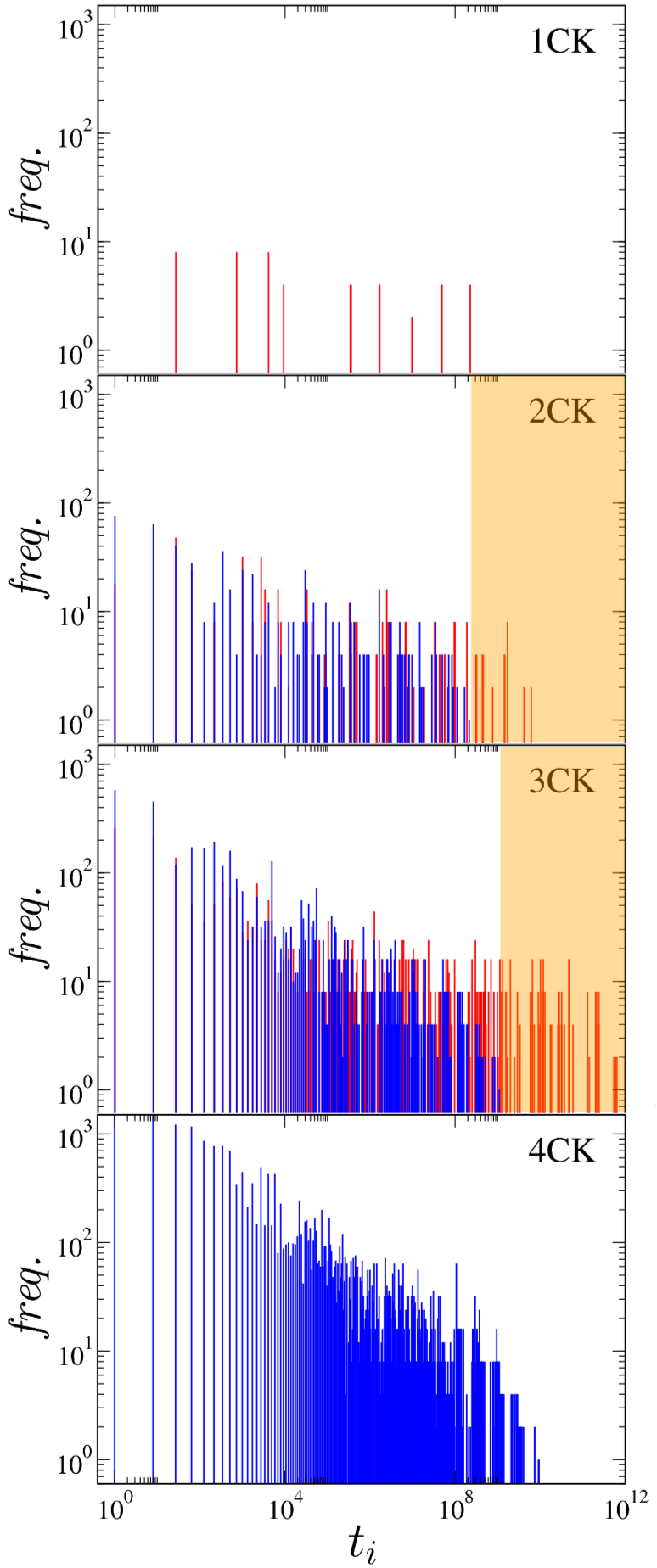}
\caption{\label{fig:poles}
Distribution of subspace block-diagonalization times, $t_i$ (in arbitrary units), comparing standard NRG (red bars) to interleaved NRG (blue bars) for MCK models. The total diagonalization time for a single NRG iteration is the sum of subspace times. The interleaved method avoids diagonalization of very large subspaces encountered in standard NRG (see shaded regions).  Parameters used: $(\Lambda,N_s)=(2.5,2000), (5,5000), (10,20000), (20,100000)$ for 1CK, 2CK, 3CK and 4CK. Data collected at the MCK fixed point.
}
\end{center}
\end{figure}

\begin{SCtable*}[1][H]
\begin{tabular}{|c|c|c|c|}
\hline
 & $\sum_i t_i^{2CK} / \sum_j t_j^{1CK^{\phantom{\dagger}}}$ & $\sum_i t_i^{3CK} / \sum_j t_j^{1CK}$ & $\sum_i t_i^{4CK} / \sum_j t_j^{1CK}$  \\[1ex]
\hline \hline
Standard NRG & 51 & 14615 & --- \\ \hline
Interleaved NRG & 2.3 & 24.6 & 438  \\ \hline
Standard/Interleaved & 22 & 594 & ---  \\ \hline
\multicolumn{4}{c}{~}\\ [1ex]\hline
 & $\sum_i m_i^{2CK} / \sum_j m_j^{1CK^{\phantom{\dagger}}}$ & $\sum_i m_i^{3CK} / \sum_j m_j^{1CK}$ & $\sum_i m_i^{4CK} / \sum_j m_j^{1CK}$\\[1ex]
\hline \hline
Standard NRG & 21 & 1310 & --- \\ \hline
Interleaved NRG & 1.6 & 7.9 & 241 \\ \hline
Standard/Interleaved & 13 & 166 & --- \\ \hline
\end{tabular} 
\caption{
Diagonalization time and memory requirements per iteration of the MCK problem, compared with reference 1CK calculation of comparable accuracy. Parameters as in Fig.~\ref{fig:poles}.
}
\label{tab:metrics}
\end{SCtable*}

For a given number of channels, one can reduce the required $N_s$ to some extent by working with a larger discretization parameter, $\Lambda$, since this increases the energy-scale separation down the Wilson chain (at the expense of introducing larger discretization artifacts). To obtain accurate results for dynamics here, we have therefore used, for both standard and interleaved NRG calculations, $(\Lambda,N_s)=(2.5,2000)$ for 1CK, $(5,5000)$ for 2CK, and $(10,20000)$ for 3CK. We find that $(20,100000)$ yields accurate results within interleaved NRG for 4CK (see also Sec.~II below).

\subsection{Block diagonalization}

To examine precisely how the computational cost depends on these values of $N_s$, one must take into account the block-diagonalization of the Hamiltonian. Using the $U(1)$ symmetries of the model, the $N_s$ retained states of each iteration are distributed amongst $N_b$ blocks. For the parameters above, we find that the maximum number of states in any \emph{single} block is $\simeq 300$ (independent of the number of channels), while $N_b$ increases with the number of channels.

These $N_b$ blocks are then used to construct new blocks at the next iteration. In standard NRG, this is done by adding an entire `shell' of Wilson chain orbitals simultaneously (one for each of the $N_c$ channels). The size of the new blocks generated in this process increases rapidly with the number of channels, since each added channel orbital has $4$ possible states. For the parameters above, we find that the largest block sizes generated within an iteration for standard NRG are $n_{\textit{full}}^{\textit{max}}\simeq 600, 1800, 10000$ for 1CK, 2CK and 3CK (data obtained at the MCK fixed point). This increase in block size is computationally costly, since the new blocks must be diagonalized, and stored in memory for calculating dynamics using the FDM-NRG.\\

To summarise, the increase of computational complexity with $N_c$ is due to increases in:
\begin{enumerate}
\item[(a)] The number of subspace blocks to be diagonalized
\item[(b)] The size of the subspace blocks to be diagonalized
\end{enumerate}

The interleaved method largely solves the problem of (b): by adding the $N_c$ Wilson chain orbitals \emph{one at a time}, truncating the subspace blocks between each addition, the matrices being diagonalized are significantly smaller than in standard NRG.  For the parameters above, we find that the largest block size to be diagonalized using the interleaved method is $n_{\textit{int}}^{\textit{max}}\simeq 500, 1000$ for 2CK and 3CK (and $n_{int}^{\textit{max}}\simeq 1500$ for 4CK).

The interleaved method does not address (a), which is inherent to the method and the physical problem itself. (One always needs to retain a sufficient number of subspace blocks at a given energy scale in order to capture the true physics.) But the effect of solving (b) produces significant savings in computational cost, as explained further below. 

\subsection{Quantitative analysis of the speedup}
Reducing the block sizes by the interleaved method has the advantage of reducing both CPU and memory requirements for the calculation. The CPU time taken to diagonalize a block of dimension $n$ scales as $n^3$ (for large $n$), and the computer memory needed to store the block scales as $n^2$. 

We define suitable platform-independent measures, proportional to the actual diagonalization time and memory usage, viz,
\begin{eqnarray}
\label{eq:time}
t_i = n_i^3 \qquad ; \qquad m_i = n_i^2 \;,
\end{eqnarray}
such that the sum over all subspaces gives the total resources required at a given iteration, $t=\sum_i t_i$ and $m=\sum_i m_i$. The distribution of $t_i$ at the MCK fixed point for 1CK, 2CK, 3CK and 4CK are given in Fig.~\ref{fig:poles}. Red bars correspond to the `full' calculation by standard NRG, while blue bars show the interleaved method. (The standard and interleaved methods are, by definition, identical in the 1CK case, so only the standard results are shown here. On the other hand, only the interleaved results are shown in the 4CK case, since the standard NRG calculation is computationally too expensive to run as far as the 4CK fixed point.)

It is clear from the 2CK and 3CK plots that the largest $t_i$ is much smaller for the interleaved method than the standard method (as pointed out in the previous section). The shaded regions highlight those subspaces in the standard method that are thereby `avoided' when using the interleaved method: noting the double-logarithmic axes, we find that 95.5\% of the total diagonalization time is spent on these subspaces in the standard 2CK calculation, while 99.4\% is spent on them for the standard 3CK calculation. 

A significant increase in efficiency is therefore obtained using the interleaved method: overall the 2CK calculation is $\sim 20$ times faster on the diagonalization step than standard NRG, while for 3CK it is $\sim 600$ times faster. The memory requirements are also much reduced: 85.7\% and 95.6\% of memory in the standard 2CK and 3CK calculations can be saved by using the interleaved method. Further metrics are given in Table \ref{tab:metrics} for the specific calculations shown in Fig.~\ref{fig:poles}.


\section{Accuracy of the Interleaved method}
Having demonstrated in the previous section that the interleaved method is significantly more efficient than standard NRG for multichannel problems, here we show that the error introduced by intermediate truncations is well-controlled and small. 

\begin{figure}[t]
\begin{center}
\includegraphics[width=75mm]{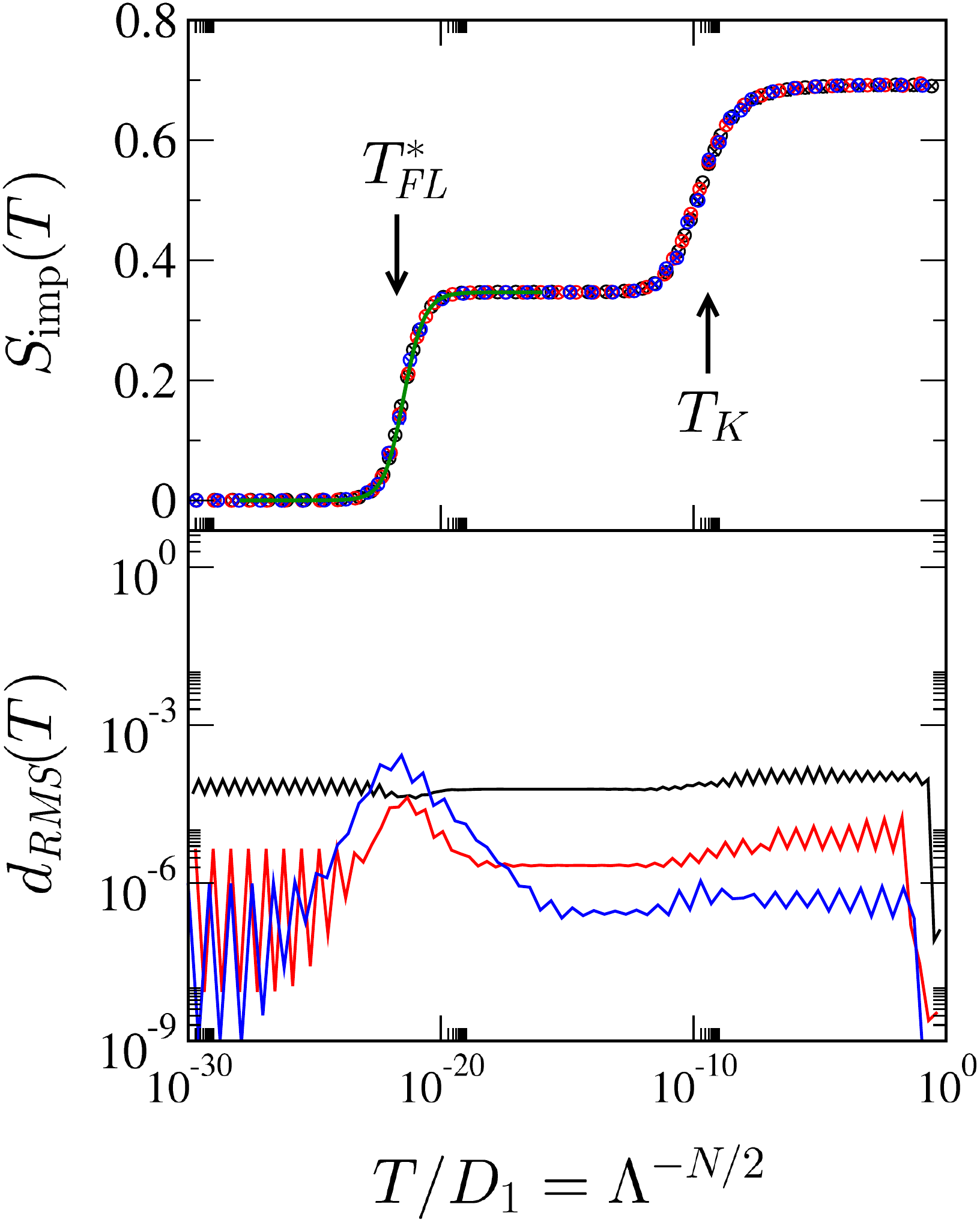}
\caption{\label{fig:relerr}
\emph{Upper panel:} Impurity contribution to entropy for the 2CK model, $S_{\text{imp}}(T)$ vs $T/D_1$ ($= \Lambda^{-N/2}$) for $\Lambda=3,5,7$ (black, red and blue colors), calculated via standard NRG (circle points) and interleaved NRG (cross points). $g=\Lambda^{-1/4}$, $N_s=5000$ and $\rho_1 J_1=0.05$ were used, with $\rho_2 J_2 \approx \rho_1 J_1$ tuned in each case to give the same crossover scale $T_{\textit{FL}}^*$. Solid line is the exact result of Eq.~\ref{entropycross} for the low-temperature crossover. \emph{Lower panel:} Error $d_{\textit{RMS}}(T)$, defined in Eq.~\ref{eq:relerr}, for the same systems.
}
\end{center}
\end{figure}

To put such an analysis on firm footing, we consider now the 2CK model, which can be solved accurately (if more laboriously) using standard NRG. Specifically, we use $\rho_1 J_1=0.05$, and retain $N_s=5000$ states at each iteration. We choose a range of discretization parameters $\Lambda$, employing for each a bandwidth ratio $g=\Lambda^{-1/4}$.

In Fig.~\ref{fig:relerr} (upper panel), we show the impurity contribution to the entropy, $S_{\text{imp}}(T)$ vs $T/D_1$ for $\Lambda=3,5$ and $7$ (black, red and blue colors). The circle points are obtained from standard NRG, while the cross points are from interleaved NRG. We have tuned $\rho_2 J_2\approx \rho_1 J_1$ in each case to generate the same Fermi liquid crossover scale, $T^*_{\textit{FL}}$. As seen clearly from the figure, all calculations coincide over the entire temperature range. As further confirmation of the accuracy of interleaved NRG, we plot as the solid line the exact result for the low-temperature crossover,\cite{gogolin,akm:finiteT}
\begin{equation}
\label{entropycross}
S_{\text{imp}}(T) \overset{T\ll T_K}{\sim} \tfrac{1}{2}\ln (2) +\bar{S}\left(a\, \frac{T}{T_{\textit{FL}}^*} \right),
\end{equation}
in terms of the universal function,
\begin{equation}
\bar{S}(t) = \frac{1}{t} \left[ \psi\left(
    \frac{1}{2}+\frac{1}{ t} \right)-1\right ]- \ln \left
  [ \frac{1}{\sqrt{\pi}}\Gamma \left( \frac{1}{2}+\frac{1}{ t}
  \right)\right ],
\end{equation}
where $\psi(z)$ is the psi (digamma) function, and $a=\mathcal{O}(1)$. 

The agreement confirms that the additional Hilbert space truncations at intermediate steps within interleaved NRG do not noticeably affect thermodynamic quantities. Note that excellent agreement between standard NRG and the interleaved method was also obtained for the t matrix in Fig.~2(a) of the main paper (dot-dashed line and circle points); see also Sec.~III, below. 

To quantify the error precisely, we consider now the root mean squared difference in energies of the eigenstates from the standard and interleaved NRG methods, calculated for all eigenstates $i$ at a given iteration $N$ and for the same $\Lambda$. We define
\begin{eqnarray}
\label{eq:relerr}
d_{\textit{RMS}}(N) = \frac{1}{N_s}\sqrt{\sum_i \left (E_{N,i}^{\textit{std}} - E_{N,i}^{\textit{int}}\right )^2 } \;,
\end{eqnarray}
which is plotted in the lower panel of Fig.~\ref{fig:relerr} for the same systems. Since the truncation is justified by energy-scale separation controlled by $\Lambda$, one naturally expects the error to decrease as $\Lambda$ increases. This is indeed found to be the case, with the characteristic error for $\Lambda=3, 5$ and $7$ being $\sim 10^{-4}, 10^{-5}$ and $10^{-6}$ respectively. The increase in $d_{\textit{RMS}}(T)$ around $T^*_{\textit{FL}}$ is simply due to the fact that the relevant perturbation generating the crossover grows under RG at slightly different rates for slightly different $T^*_{\textit{FL}}$. We stress that the fixed points and crossovers are described accurately, as is the universal critical physics which depends only on proximity to the critical point $|\rho_2 J_2 - \rho_2 J_2^c|$, not on $\rho_2 J_2$ itself. 

\begin{figure}[t]
\begin{center}
\includegraphics[width=75mm]{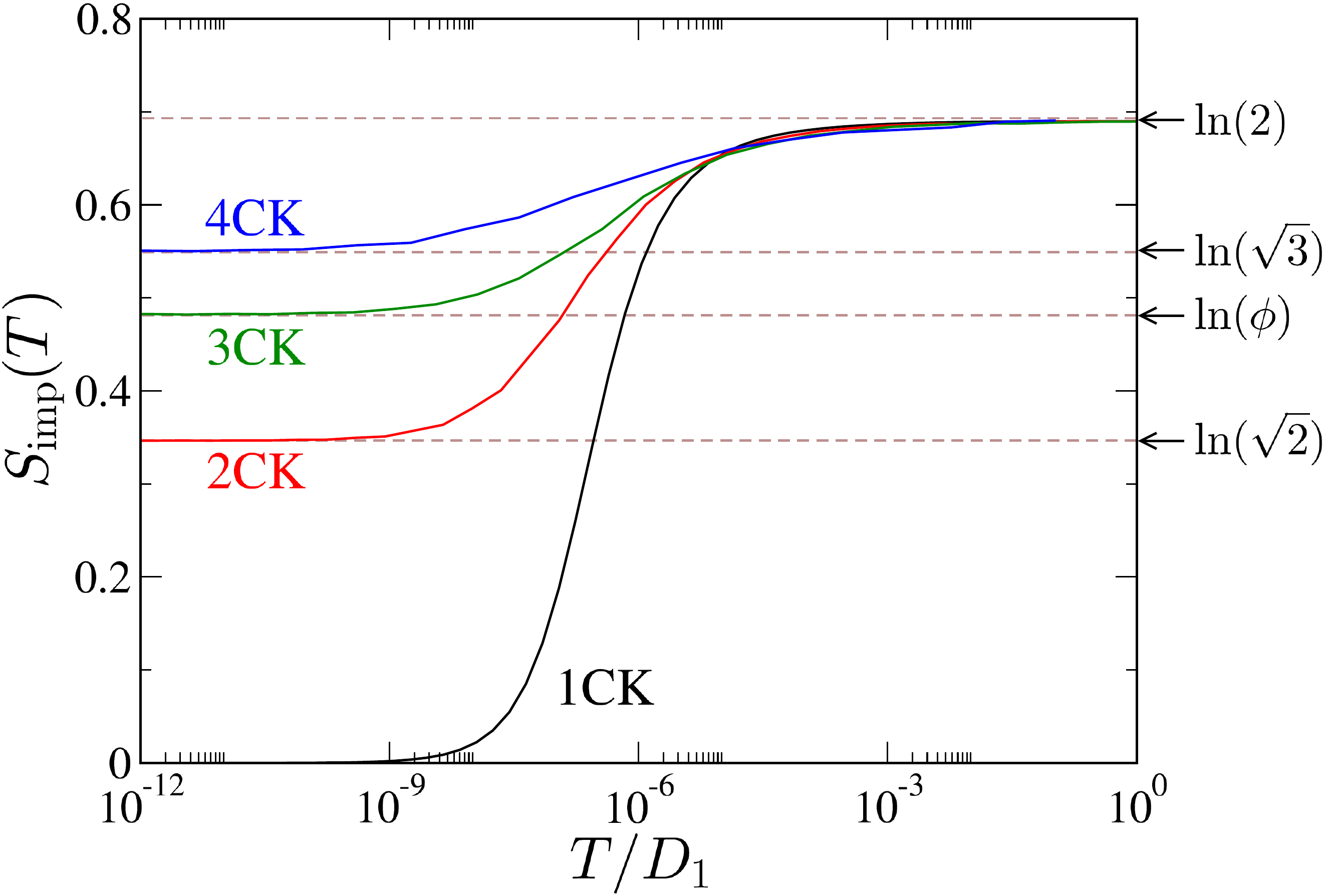}
\caption{\label{fig:mck_ent}
Comparison of entropy $S_{\text{imp}}(T)$ vs $T/D_1$ for 1CK, 2CK, 3CK and 4CK models, calculated using interleaved NRG. The same parameters as Fig.~\ref{fig:poles} were used, and $\rho_1 J_1 = 0.075$ with $\rho_2 J_2\approx \rho_1 J_1$ tuned to the MCK critical point in each case. Results with $z=0$ and $0.5$ were averaged for 3CK and 4CK to reduce discretization artifacts. The 4CK critical point was located using calculations in which $N_s=65000$ states were kept. $\phi=\tfrac{1}{2}[1+\sqrt{5}]$ is the golden ratio. 
}
\end{center}
\end{figure}

To emphasize that the MCK critical point can be captured by interleaved NRG, and to demonstrate the method in action for the 4CK model, we show in Fig.~\ref{fig:mck_ent} a comparison of the impurity entropy for 1CK, 2CK, 3CK and 4CK models for common $\rho_1 J_1 = 0.075$ and $\rho_2 J_2\approx \rho_1 J_1$ tuned to the MCK critical point in each case. The MCK fixed point entropies reproduce the known results of Ref.~\onlinecite{andreidestri,tsvelik}.


\section{Emergent channel symmetry}
As shown above and in the main paper, the NFL critical physics of the MCK model is still realized when the conduction electron channels have different bandwidths. This is because the critical physics can be accessed by \emph{tuning} the coupling constants of the model, to `cancel out' the symmetry breaking induced by the different bandwidths. An emergent channel symmetry then arises. We explain this point in more detail below.

In the wide, flat-band limit $D_1\rightarrow \infty$ it is clear that the quantum critical point of the MCK model (with $N_c>1$ channels) arises when $\rho_{\alpha} J_{\alpha} = \rho_{\beta} J_{\beta}$ for any fixed finite ratio $g=D_{\alpha}/D_{\beta}$: all conduction bands are infinitely wide, so the parameter $g$ `drops out' and the coupling of the impurity to channel $\alpha$ is described completely by the dimensionless coupling strength $\rho_{\alpha} J_{\alpha}$. \cite{nozieres} The crossover from the local moment fixed point to the NFL fixed point is then universal, and is the same for all $g$. 

For finite bandwidths, the same universal curve is obtained, but only on universal energy scales $\ll J_{\alpha}$, $D_{\alpha}$. This is well known in the case of equal bandwidths ($g=1$): one argues that the RG flows on energy scales much less than the bandwidth are the same as those in the infinite bandwidth case, the non-universal physics having been integrated out. 
\emph{We point out that the same is true for $g\ne 1$}. Of course, at high energies the channel symmetry breaking then causes each dimensionless coupling constant $\rho_{\alpha} J_{\alpha}$ to evolve differently under RG. But on the \emph{universal} energy scales below the smallest of the bandwidths, the scaling under RG becomes the same as that of the infinite-bandwidth case, and hence the same universal physics---including the the same quantum critical behavior---is expected. 

To access the critical physics, one must simply choose the  bare $\rho_{\alpha} J_{\alpha}$ such that the coupling to each lead is identical \emph{once the universal scaling regime is reached}. For $g=1$, this requires that the bare $\rho_{\alpha} J_{\alpha}$ are equal by symmetry. For $g\ne 1$, since the couplings scale differently at high energies, one must start from different bare couplings so that the effective couplings become equal in the universal regime. In other words, the bare couplings must be tuned to locate the critical point. We stress that this is a physical feature of the model with unequal bandwidths, and not related to the interleaved NRG method itself.

To confirm this RG argument, Fig.~2(a) of the main paper shows the universal t matrix of the 2CK model calculated using interleaved NRG, with a bandwidth ratio $g=\Lambda^{-1/4}$ (and $\Lambda=5$), compared with results of standard NRG using $g=1$. The universal curves are manifestly identical, even though the bandwidths in the two cases are different.\\

[It should be pointed out that the interleaved method (relying as it does on a channel-asymmetric truncation scheme) naturally introduces a very small, additional shift in the critical $\rho_{\alpha} J_{\alpha}$. This can be seen by comparing results from the standard and interleaved methods with the same bandwidths. For the 2CK calculation with $\rho_1 J_1 = 0.05$, $\Lambda=5$ and $g=\Lambda^{-1/4}$, the tuned couplings for interleaved and standard NRG satisfy $\rho_2 J^{\textit{int}}_2 = \rho_2 J^{\textit{std}}_2 + \delta$, with the difference $\delta = 10^{-7}$. That the difference is so small suggests that \emph{once the parameters are tuned} additional symmetry-breaking due to the asymmetric truncation scheme is negligible. This conclusion is backed up by the excellent agreement between the universal scaling curves of the standard and interleaved methods. It is also seen explicitly below in results for the t matrices.]\\ 

\begin{figure}[t]
\begin{center}
\includegraphics[width=75mm]{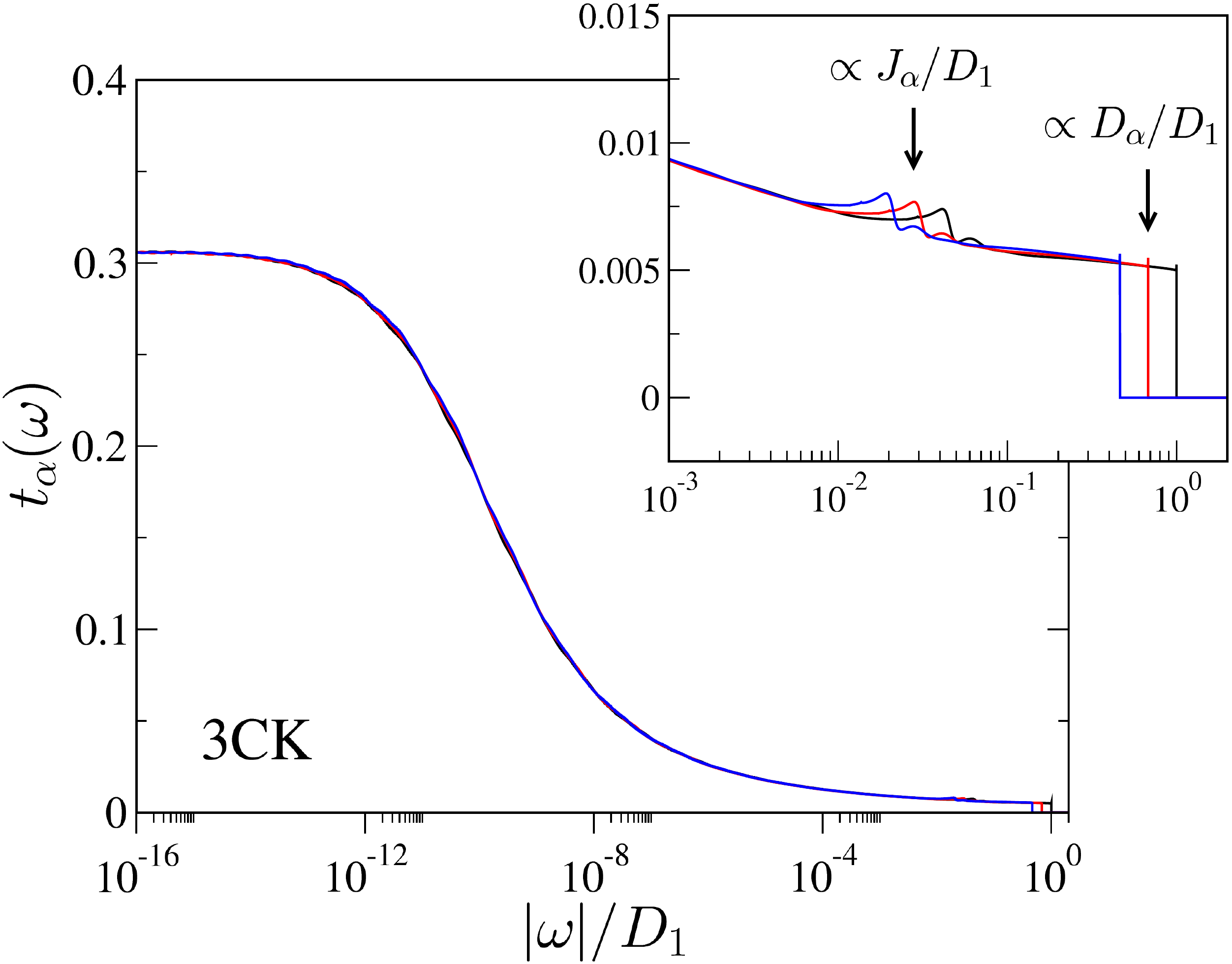}
\caption{\label{fig:tmatrices}
Spectrum of the t matrix $t_{\alpha}(\omega)$ vs $|\omega|/D_1$ for the 3CK model, with $\alpha=1,2,3$ as the black, red and blue lines respectively. Parameters used are the same as for Fig.~2(a) of the main paper. Inset shows a magnified view of non-universal high-energy features.
}
\end{center}
\end{figure}

 In the main panel of Fig.~\ref{fig:tmatrices}, we plot the spectrum of the t matrix, $t_{\alpha}(\omega)$ for each channel in the 3CK problem, as a full function of frequency $|\omega|/D_1$, after tuning to the critical point. The same parameters are used as in Fig.~2(a) of the main paper. 
The three curves are essentially indistinguishable below an energy scale $|\omega|/D_1\sim 10^{-3}$, demonstrating explicitly that the \emph{universal} RG flow in each channel is the same (i.e. there is an emergent channel symmetry), even in the case where the bandwidth ratio $g\ne 1$. 

Differences in the t matrices of the three channels arise on high, non-universal energy scales, $D_{\alpha}$ and $J_{\alpha}$, as shown in a magnified view in the inset. It is worth pointing out that the features on the scale of $J_{\alpha}$ are a real characteristic of the t matrix for a Kondo model, rather than an artifact of the interleaved NRG method itself. 

